\documentclass[aps,prd,onecolumn,groupedaddress,showpacs,nofootinbib,amssymb]{revtex4}
\usepackage[dvips]{graphicx}
\usepackage{amssymb}
\usepackage{amsmath}
\usepackage{graphicx,,color}
\usepackage{amsfonts}
\usepackage{bm}
\usepackage{cancel}
\usepackage{comment}
\newcommand\be{\begin{equation}}
\newcommand\ee{\end{equation}}

\allowdisplaybreaks[4]

\begin{document}

\title{Aspects of Axion $F(R)$ Gravity}
%\title{Inflation and Eschatological Scenarios with $f(R)$ Gravity and Axion Dark Matter}
\author{S.D. Odintsov,$^{1,2}$\,\thanks{odintsov@ieec.uab.es}
V.K. Oikonomou,$^{3,4,5}$\,\thanks{v.k.oikonomou1979@gmail.com}}
\affiliation{$^{1)}$ ICREA, Passeig Luis Companys, 23, 08010 Barcelona, Spain\\
$^{2)}$ Institute of Space Sciences (IEEC-CSIC) C. Can Magrans s/n,
08193 Barcelona, Spain\\
$^{3)}$Department of Physics, Aristotle University of Thessaloniki, Thessaloniki 54124, Greece\\
$^{4)}$ Laboratory for Theoretical Cosmology, Tomsk State
University of Control Systems
and Radioelectronics, 634050 Tomsk, Russia (TUSUR)\\
$^{5)}$ Tomsk State Pedagogical University, 634061 Tomsk, Russia }

\tolerance=5000

\begin{abstract}
We provide a compact review on recent developments on axion $F(R)$
gravity. The axion field is a string theory originating
theoretical particle that is a perfect candidate for low-mass
particle dark matter. In this review we present how a viable
inflationary phenomenology and a viable late-time evolution can be
described by an axion $F(R)$ gravity theory, in which the $F(R)$
gravity part can drive in a geometric way the inflationary and the
late-time era, and the axion field behaves as dark matter, with
its energy density $\rho_a$ behaving as a function of the scale
factor as $\rho_a\sim a^{-3}$. We also briefly discuss the effect
of a non-trivial axion Chern-Simons coupling on the inflationary
phenomenology of the $R^2$ model. Finally, we briefly discuss the
effects of a non-minimal coupling of the axion field with the
curvature on neutron stars, and also the propagation of gravity
waves in Chern-Simons axion gravity.
\end{abstract}

%PACS numbers: 04.50.Kd, 95.36.+x, 98.80.-k, 98.80.Cq

\maketitle

\section{Introduction}

Dark matter and dark energy are two of the most persisting
problems in theoretical physics and theoretical cosmology. Dark
energy is a name that characterizes the still unknown mechanism
that drives the acceleration of the Universe, firstly observed in
the late 90's \cite{Riess:1998cb}. This observational result was
one of the most surprising phenomena ever observed, and the whole
problem becomes even more difficult at present time, since
observational data indicate that the expansion rate of the
Universe based on low redshift sources is different in comparison
to the data obtained by the Cosmic Microwave Background radiation
\cite{Aghanim:2018eyx}. This problem is nowadays known as the
$H_0$ tension problem \cite{Riess:2011yx,Busti:2014dua}, and there
exist a plethora of data coming from different sources that verify
this tension in the exact value of the Hubble rate. With regard to
dark matter, the emergence of the dark matter concept appeared
after the first galactic rotation curves were studied, and its
existence is a theoretical proposal even up to date, since no dark
matter particle has ever been found. The experiments for nearly
three decades focused on finding a Weakly Interacting Massive
Particle (WIMP) with mass MeV or even hundreds GeV with no
results. For reviews and some characteristic articles related to
the heavy mass dark matter particle searches can be found in Refs.
\cite{Bertone:2004pz,Bergstrom:2000pn,Mambrini:2015sia,Profumo:2013yn,Hooper:2007qk,Oikonomou:2006mh}.
The last five years, experiments and observations are focused on
finding small mass WIMP, in the mass range of eV or even much
smaller than the eV scale. The most important small mass WIMP
candidate is the axion
\cite{Marsh:2015xka,Sikivie:2006ni,Raffelt:2006cw,Linde:1991km},
which is a string theory and quantum chromodynamics (QCD)
originating particle. The QCD axion is accompanied by many
theoretical problems, since the primordial $U(1)$ Peccei-Quinn
symmetry is not broken during inflation, and thus the most
plausible axion scenario for viable model building is the
misalignment axion scenario \cite{Marsh:2015xka}. In the
literature there exist quite many articles focusing on the axion
dark matter perspective, see for example
\cite{Marsh:2015xka,Marsh:2017yvc,Co:2019jts,Odintsov:2019mlf,Nojiri:2019nar,Nojiri:2019riz,Odintsov:2019evb,Cicoli:2019ulk,Fukunaga:2019unq,Caputo:2019joi,maxim,Chang:2018rso,Irastorza:2018dyq,Anastassopoulos:2017ftl,Sikivie:2014lha,Sikivie:2010bq,Sikivie:2009qn,Caputo:2019tms,Masaki:2019ggg,Aoki:2017ehb,Aoki:2016kwl,Obata:2016xcr,Aoki:2016mtn,Ikeda:2019fvj,Arvanitaki:2019rax,Arvanitaki:2016qwi,Arvanitaki:2014wva,Arvanitaki:2014dfa,Sen:2018cjt,Cardoso:2018tly,Rosa:2017ury,Yoshino:2013ofa,Machado:2019xuc,Korochkin:2019qpe,Chou:2019enw,Chang:2019tvx,Crisosto:2019fcj,Choi:2019jwx,Kavic:2019cgk,Blas:2019qqp,Guerra:2019srj,Tenkanen:2019xzn,Huang:2019rmc,Croon:2019iuh,Day:2019bbh,Odintsov:2020nwm}.
In addition to the theoretical works, there exist additional
experimental and theoretical proposals based on the assumption of
the existence of axion particle dark matter
\cite{Du:2018uak,Henning:2018ogd,Ouellet:2018beu,Safdi:2018oeu,Rozner:2019gba,Avignone:2018zpw,Caputo:2018vmy,Caputo:2018ljp,Lawson:2019brd}.
In the majority of the experimental proposals, the axion-photon
conversion in a magnetic field feature is utilized
\cite{Balakin:2009rg,Balakin:2012up,Balakin:2014oya}. A
particularly interesting approach to low mass scalar dark matter
particles is the effect of the latter on gravity waves coming from
astrophysical sources
\cite{Arvanitaki:2019rax,Arvanitaki:2016qwi,Arvanitaki:2014wva} or
even from primordial gravitational waves \cite{Satoh:2007gn},
where the possibility of finding non-trivial polarizations is
mainly studied or stressed
\cite{Inomata:2018rin,Kamionkowski:2000gb}. The non-trivial
polarization of gravitational waves in theories involving
Chern-Simons axionic couplings is also discussed in the literature
\cite{Satoh:2007gn}. For an important stream of paper on
Chern-Simons coupling terms see
\cite{Nishizawa:2018srh,Wagle:2018tyk,Yagi:2012vf,Yagi:2012ya,Molina:2010fb,Izaurieta:2009hz,Sopuerta:2009iy,Konno:2009kg,Smith:2007jm,Matschull:1999he,Haghani:2017yjk,Satoh:2007gn,Satoh:2008ck,Yoshida:2017cjl,Nojiri:2019nar}.
Moreover, the axions can be related with the $H_0$-tension problem
\cite{DEramo:2018vss}.

Modified gravity
\cite{reviews1,reviews2,reviews3,reviews4,reviews5,reviews6} can
successfully describe in a viable way in many cases, the dark
energy and dark matter problems. For an important stream of
articles on the modified gravity descriptions of the dark sector
of our Universe, see for example
\cite{Capozziello:2002rd,Carroll:2003wy,Nojiri:2003ft,Nojiri:2007as,Nojiri:2007cq,Cognola:2007zu,Nojiri:2006gh,Benetti:2019gmo,Nojiri:2008nk,Anagnostopoulos:2019myt,Benisty:2020nuu}.
Although certain dynamical aspects of dark matter can be explained
by modified gravity \cite{Capozziello:2006uv,Capozziello:2012ie},
observational data coming from the Bullet cluster indicate that
dark matter might have eventually a particle nature. To our
opinion, unless supersymmetry is found in the Large Hadron
Collider, which may offer new insights in the particle dark matter
problem, the axion is particle dark matter's last hope.

In this short review we shall present the latest developments on
axion $F(R)$ gravity phenomenology. Particularly, we shall present
several models of $F(R)$ gravity in the presence of a misalignment
axion scalar field, and we shall demonstrate how a viable
inflationary phenomenology can be achieved. In the simplest case,
the axion scalar and the $F(R)$ gravity are decoupled, and due to
the fact that at early times the misalignment axion is frozen in
its primordial vacuum expectation value, the inflationary era is
controlled by the $R^2$ part of the $F(R)$ gravity, and as the
Hubble rate drops and becomes approximately equal to the axion
scalar mass, the axion begins to oscillate, in a way so that its
energy density $\rho_a$ scales as $\rho_a\sim a^{-3}$. Thus, the
axion field behaves perfectly as dark matter, and this behavior
continues until late times. The $F(R)$ gravity contains terms
which also affect the late-time era, and thus we will show that a
viable phenomenology, compatible with the $\Lambda$CDM model in
most of the cases, is obtained. Qualitatively similar results are
obtained for a model with a non-minimal coupling between the axion
and the one of the $F(R)$ gravity components. Moreover, we discuss
how a Chern-Simons coupling can reduce the tensor-to-scalar ratio
of the Starobinsky model \cite{Starobinsky:1980te} to a great
extent. Finally, we demonstrate how a non-minimal coupling between
the scalar axion and the curvature can affect the maximum mass of
a neutron star, while leaving the radius unaffected.

\section{Misalignment Axion-$F(R)$ Gravity Model With No and with non-Minimal Coupling}

\subsection{Model I}

In this subsection we analyze the first axion $F(R)$ model, the
details on this model can be found in Ref.
\cite{Odintsov:2020nwm}. The first axion-$F(R)$ gravity model we
shall present, has the following action
\cite{Odintsov:2019evb,Odintsov:2020nwm}
\begin{equation}
\label{mainaction} \mathcal{S}=\int d^4x\sqrt{-g}\left[
\frac{1}{2\kappa^2}F(R)-\frac{1}{2}\partial^{\mu}\phi\partial_{\mu}\phi-V(\phi)+\mathcal{L}_m
\right]\, ,
\end{equation}
with $\kappa^2=\frac{1}{8\pi G}=\frac{1}{M_p^2}$, and $G$ is
Newton's gravitational constant while $M_p$ is the reduced Planck
mass. The Lagrangian $\mathcal{L}_m$ represents all the perfect
matter fluids present. The functional form of the $F(R)$ gravity
has the following form,
\begin{equation}\label{starobinsky}
F(R)=R+\frac{1}{M^2}R^2-\gamma \Lambda
\Big{(}\frac{R}{3m_s^2}\Big{)}^{\delta}\, ,
\end{equation}
In addition $m_s$ in Eq. (\ref{starobinsky}) stands for
$m_s^2=\frac{\kappa^2\rho_m^{(0)}}{3}$, and we choose the
parameter $\delta $ to be in the interval $0<\delta <1$. For
phenomenological reasons, the parameter $M$ is $M= 1.5\times
10^{-5}\left(\frac{N}{50}\right)^{-1}M_p$, with $N$ standing for
the $e$-foldings number. For a flat Friedmann-Robertson-Walker
(FRW) metric,
\begin{equation}
\label{metricfrw} ds^2 = - dt^2 + a(t)^2 \sum_{i=1,2,3}
\left(dx^i\right)^2\, ,
\end{equation}
the equations of motion are,
\begin{align}\label{eqnsofmkotion}
& 3 H^2F_R=\frac{RF_R-F}{2}-3H\dot{F}_R+\kappa^2\left(
\rho_r+\frac{1}{2}\dot{\phi}^2+V(\phi)\right)\, ,\\ \notag &
-2\dot{H}F_R=\kappa^2\dot{\phi}^2+\ddot{F}_R-H\dot{F}_R
+\frac{4\kappa^2}{3}\rho_r\, ,
\end{align}
\begin{equation}\label{scalareqnofmotion}
\ddot{\phi}+3H\dot{\phi}+V'(\phi)=0
\end{equation}
with $F_R=\frac{\partial F}{\partial R}$. Moreover the ``dot'' and
the ``prime'' denote differentiation with respect to the cosmic
time and the scalar field respectively, and also the only matter
fluid present is that of radiation  $p_r=\frac{1}{3}\rho_r$. As we
show in the next section, the axion scalar field will act as the
dark matter perfect fluid.

The dynamical evolution of the misalignment axion field is crucial
for all the axion $F(R)$ gravity models we shall develop. The
scalar field potential of the misalignment axion after the
primordial $U(1)$ Peccei-Quinn symmetry was broken, has the form
$V(\phi )\simeq \frac{1}{2}m_a^2\phi^2_i$, where $\phi_i$ is the
vacuum expectation value of the axion field after the $U(1)$
symmetry is broken. During the early times when inflation occurs,
the dynamical evolution of the axion is overdamped, and satisfies
the initial conditions $\ddot{\phi}(t_i)\simeq 0$,
$\dot{\phi}(t_i)\simeq 0$ and $ \phi(t_i)\equiv\phi_i=f_a\theta_a$
\cite{Marsh:2015xka}, where $t_i$ is a characteristic time scale
during inflation, and $f_a$ is the axion decay constant, while
$\theta_a$ is the misalignment angle. The axion remains frozen as
long as $H\gg m_a$, however, when $H\sim m_a$, axion oscillations
occur. Assuming a slowly varying evolution, it can be shown that
the axion energy density scales as $\rho_a\sim a^{-3}$
\cite{Marsh:2015xka,Odintsov:2020nwm}.

\subsubsection{The Inflationary Era: $R^2$ Gravity Domination}

Now let us show that indeed during the inflationary era, the axion
field has no effect on the dynamics. The equations of motion
(\ref{eqnsofmkotion}) during inflation are,
\begin{equation}\label{friedmanequationinflation}
3H^2\left(1+\frac{2}{M^2}R-\delta \gamma
\Big{(}\frac{R}{3m_s^2}\Big{)}^{\delta-1}\right)=\frac{R^2}{2M}+(\gamma-\gamma
\delta
)\frac{\Big{(}\frac{R}{3m_s^2}\Big{)}^{\delta}}{2}-3H\dot{R}\Big{(}\frac{2}{M^2}-\gamma
\delta (\delta-1)\Big{(}\frac{R}{3m_s^2}\Big{)}^{\delta-2}\Big{)}+
\kappa^2\Big{(}\rho_r+\frac{1}{2}\kappa^2\dot{\phi_i}^2+\frac{1}{2}m_a^2\phi_i^2
\Big{)}\, .
\end{equation}
For late-time phenomenological reasons, we choose,
\begin{equation}\label{gammaanddelta}
\gamma=\frac{1}{0.5},\,\,\, \delta=\frac{1}{100}\, ,
\end{equation}
and $\Lambda\simeq 11.895\times 10^{-67}$eV$^2$. Also for $N\sim
60$, $M$ is $M\simeq 3.04375\times 10^{22}$eV. In addition, during
inflation, $H= H_I\sim 10^{13}$GeV so $R\sim 1.2\times
10^{45}$eV$^2$. The radiation fluid term is  $\kappa^2\rho_r \sim
e^{-N}$ so it can safely be ignored,  and also since
$\phi_i=\mathcal{O}(10^{15})$GeV we have $m_a\simeq
 \mathcal{O}(10^{-14})$eV. In effect, the potential term is $\kappa^2V(\phi_i)\sim \mathcal{O}(8.41897\times
 10^{-36})$eV$^{2}$. The curvature scalar related terms are of the
 order$R\sim 1.2\times \mathcal{O}(10^{45})$eV$^2$, and $R^2/M^2\sim \mathcal{O}(1.55\times
 10^{45})$eV$^2$, and in addition, $\sim \Big{(}\frac{R}{3m_s^2}\Big{)}^{\delta}\sim
 \mathcal{O}(10)$,  and also $\sim \Big{(}\frac{R}{3m_s^2}\Big{)}^{\delta-1}\sim
 \mathcal{O}(10^{-111})$ and finally $\sim \Big{(}\frac{R}{3m_s^2}\Big{)}^{\delta-2}\sim
 \mathcal{O}(10^{-223})$. From the above it is evident that the
 Ricci scalar related terms dominate the inflationary evolution,
 and specifically the $R^2$ term, hence the model is reduced to
 the $R^2$ model, which yields a viable inflationary phenomenology
 compatible with the observational data coming from Planck 2018 \cite{Aghanim:2018eyx}.

\subsubsection{Dark Energy Era}

\begin{figure}[h!]
\centering
\includegraphics[width=18pc]{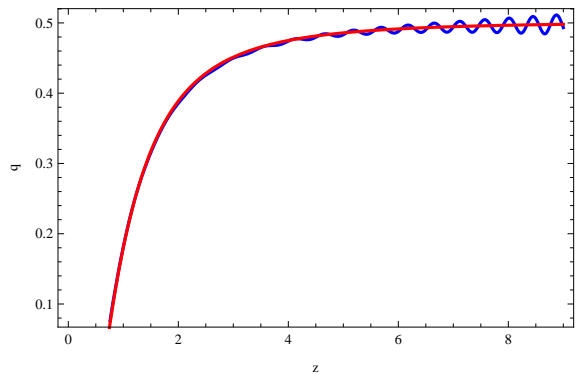}
\caption{The deceleration parameter $q$ of the axion $F(R)$
gravity model (blue curve) and of the $\Lambda$CDM model  for
$\tilde{\gamma}=1/10^3$ for $z=[0,9]$.} \label{plot1}
\end{figure}

Let us now consider the late-time era, and the equations of motion
can be cast in the Einstein field equations form for a FRW metric
as follows,
\begin{align}\label{flat}
& 3H^2=\kappa^2\rho_{tot}\, ,\\ \notag &
-2\dot{H}=\kappa^2(\rho_{tot}+P_{tot})\, ,
\end{align}
with $\rho_{tot}=\rho_{\phi}+\rho_G+\rho_r$ being the total energy
density and $P_{tot}=P_r+P_{a}+P_{G}$ being the total pressure of
the cosmological fluid consisting of the axion scalar field, the
radiation perfect fluid and the geometric contribution coming from
the $F(R)$ gravity. In our case $\rho_G$ and $P_G$ are equal to,
\begin{equation}\label{degeometricfluid}
\kappa^2\rho_{G}=\frac{F_R R-F}{2}+3H^2(1-F_R)-3H\dot{F}_R\, ,
\end{equation}
\begin{equation}\label{pressuregeometry}
\kappa^2P_G=\ddot{F}_R-H\dot{F}_R+2\dot{H}(F_R-1)-\rho_G\, .
\end{equation}
Using the redshift $z$ as a dynamical variable,and by introducing
the statefinder quantity $y_H(z)$ \cite{Hu:2007nk,Bamba:2012qi},
\begin{equation}\label{yHdefinition}
y_H(z)=\frac{\rho_{G}}{\rho^{(0)}_m}\, ,
\end{equation}
with $\rho^{(0)}_m$ being the cold dark matter energy density at
present time. We can write $y_H(z)$ in terms of the Friedmann
equation (\ref{flat}),
\begin{equation}\label{yhfunctionanalyticzero}
y_H(z)=\frac{3H^2}{\kappa^2\rho^{(0)}_m}-\frac{\dot{\phi}^2}{2\rho^{(0)}_m}-\frac{V(\phi)}{\rho^{(0)}_m}-\frac{\rho_r}{\rho^{(0)}_m}\,
.
\end{equation}
With regard to the radiation energy density this scales as
$\rho_r=\rho_r^{(0)}a^{-4}$, with $\rho_r^{(0)}$ being the value
of the radiation energy density at  present time , hence
$\frac{\rho_r}{\rho^{(0)}_m}=\chi (1+z)^4$, with
$\chi=\frac{\rho^{(0)}_r}{\rho^{(0)}_m}\simeq 3.1\times 10^{-4}$.
We can write Eq. (\ref{yhfunctionanalyticzero}) as follows,
\begin{equation}\label{finalexpressionyHz}
y_H(z)=\frac{H^2}{m_s^2}-(1+z)^{3}-\chi (1+z)^4\, .
\end{equation}
with
$m_s^2=\frac{\kappa^2\rho^{(0)}_m}{3}=H_0\Omega_c=1.37201\times
10^{-67}$eV$^2$. The Friedmann equation of motion can be cast in
terms of the statefinder function $y_H(z)$ as follows
\cite{Bamba:2012qi},
\begin{equation}\label{differentialequationmain}
\frac{d^2y_H(z)}{d z^2}+J_1\frac{d y_H(z)}{d z}+J_2y_H(z)+J_3=0\,
,
\end{equation}
with the functions $J_1$, $J_2$ and $J_3$ being defined in the
following way,
\begin{align}\label{diffequation}
& J_1=\frac{1}{z+1}\left(
-3-\frac{1-F_R}{\left(y_H(z)+(z+1)^3+\chi (1+z)^4\right) 6
m_s^2F_{RR}} \right)\, , \\ \notag & J_2=\frac{1}{(z+1)^2}\left(
\frac{2-F_R}{\left(y_H(z)+(z+1)^3+\chi (1+z)^4\right) 3
m_s^2F_{RR}} \right)\, ,\\ \notag & J_3=-3(z+1)-\frac{\left(1-F_R
\right)\Big{(}(z+1)^3+2\chi (1+z)^4
\Big{)}+\frac{R-F}{3m_s^2}}{(1+z)^2\Big{(}y_H(z)+(1+z)^3+\chi(1+z)^4\Big{)}6m_s^2F_{RR}}\,
,
\end{align}
with $F_{RR}=\frac{\partial^2 F}{\partial R^2}$. We shall solve
numerically the above equation, focusing on the redshift interval
$z=[z_i,z_f]$ with $z_i=0$ and $z_f=10$. In addition we shall use
the following initial conditions at redshift $z_f=10$,
\begin{equation}\label{generalinitialconditions}
y_H(z_f)=\frac{\Lambda}{3m_s^2}\left(
1+\tilde{\gamma}(1+z_f)\right)\, , \,\,\,\frac{d y_H(z)}{d
z}\Big{|}_{z=z_f}=\tilde{\gamma}\frac{\Lambda}{3m_s^2}\, ,
\end{equation}
where $\tilde{\gamma}$ is $\tilde{\gamma}=\frac{1}{10^3}$. Using
the numerical solution for $y_H(z)$ we can compare several
statefinder quantities of the axion $F(R)$ model with the
$\Lambda$CDM model. Moreover $\Omega_r/\Omega_M\simeq \chi$, where
$\chi$ is defined below Eq. (\ref{yhfunctionanalyticzero}). In the
literature many statefinder quantities are studied
\cite{Sahni:2014ooa}, however here we shall consider only the
deceleration parameter $q=-1-\frac{\dot{H}}{H^2}$. In Fig.
\ref{plot1} we present the comparison of the deceleration
parameter $q$ as a function of the redshift for the axion $F(R)$
gravity model (blue curve) and for the $\Lambda$CDM model (red
curve), for $z=[0,9]$. As it can be seen the models are
indistinguishable, and only for large redshifts $z>6$ the
deceleration parameter of the axion $F(R)$ gravity models shows
some oscillating behavior.

\subsection{Model II}

The second axion $F(R)$ gravity model which we shall consider has
the following gravitational action \cite{Odintsov:2019evb},
\begin{equation}
\label{mainaction} \mathcal{S}=\int d^4x\sqrt{-g}\left[
\frac{1}{2\kappa^2}f(R)+\frac{1}{2\kappa^2}h(\phi)G(R)-\frac{1}{2}\partial^{\mu}\phi\partial_{\mu}\phi-V(\phi)
\right]\, ,
\end{equation}
We introduce the notation for simplicity,
\begin{equation}\label{gravisetnotation}
\mathcal{F}(R,\phi)=\frac{1}{\kappa^2}f(R)+\frac{1}{\kappa^2}h(\phi)G(R)\,
.
\end{equation}
Varying the gravitational action with respect to the metric we
get,
\begin{align}\label{eqnsofmkotion}
& 3 H^2F=\frac{1}{2}\dot{\phi}^2+\frac{RF-\mathcal{F}+2
V}{2}-3H\dot{F}\, ,\\ \notag &
-3FH^2-2\dot{H}F=\frac{1}{2}\dot{\phi}^2-\frac{RF-\mathcal{F}+2
V}{2}+\ddot{F}+2H\dot{F}\, ,
\end{align}
with $F=\frac{\partial \mathcal{F}}{\partial R}$, and the
variation with respect to the scalar field is,
\begin{equation}\label{scalarfieldeqn}
\ddot{\phi}+3H\dot{\phi}+\frac{1}{2}\left(-\mathcal{F}'(R,\phi)+2V'(\phi)\right)=0\,
.
\end{equation}
We choose the $f(R)$ gravity to be the same as in the previous
section, and for late-time phenomenological reasoning, we choose
$G(R)$ to be,
\begin{equation}\label{GRfunction}
G(R)\sim R^{\gamma}\, ,
\end{equation}
with $\gamma$ taking values $0<\gamma <0.75$. The early time era
behavior of the model is similar to the model presented in the
previous section, since the $R^2$ model dominates the evolution of
the Universe \cite{Odintsov:2019evb}. At late-times the $\sim R^2$
terms are subdominant, thus the Friedmann equation becomes,
\begin{equation}\label{friedmanequationinflationdarkenergy}
3H^2\left(1+h(\phi)G'(R)\right)\simeq
\frac{1}{2}\kappa^2\dot{\phi}^2+2V\kappa^2-3H\dot{R}G''(R)h(\phi)\,
.
\end{equation}
Thus at late times, the dominant terms in the left hand side of
Eq. (\ref{friedmanequationinflationdarkenergy}) are $\sim
h(\phi)G'(R)$ since $G'(R)\sim R^{\gamma-1}$ due to the fact that
$0<\gamma<0.75$. On the other hand, in the right hand side of Eq.
(\ref{friedmanequationinflationdarkenergy}), the term
$-3H\dot{R}G''(R)h(\phi)$ dominates, since it contains powers of
the curvature of the form $G''(R)\sim R^{\gamma-2}$. In effect,
the leading order form of the Friedmann equation reads
\cite{Odintsov:2019evb},
\begin{equation}\label{finalfriedmanequationlatetimes}
RH\simeq \left(1-\gamma \right)\dot{R}\, .
\end{equation}
By using $R=12H^2+6\dot{H}$ the solution of the resulting
differential equation at leading order is approximately,
\begin{equation}\label{latetimedesitterevolutin}
H(t_0)\simeq \frac{\sqrt{2} \sqrt{1-\gamma } \sqrt{\Lambda
}}{\sqrt{3-4 \gamma }}\, ,
\end{equation}
so we obtain a de Sitter late-time evolution. Hence it is obvious
that the axion-$f(R)$ gravity non-minimal coupling controls the
late-time era.

\subsection{Effect of Chern-Simons Axion Terms on the $R^2$ Model}

In this section we shall demonstrate how a Chern-Simons coupling
can reduce the tensor-to-scalar ratio of the standard $R^2$ model.
The action is \cite{Odintsov:2019mlf},
\begin{equation}
\label{mainactionreview} \mathcal{S}=\int d^4x\sqrt{-g}\left[
\frac{1}{2\kappa^2}f(R)-\frac{1}{2}\partial^{\mu}\phi\partial_{\mu}\phi-V(\phi)+\frac{1}{8}\nu
(\phi)R\tilde{R} \right]\, ,
\end{equation}
with $R\tilde{R}=\epsilon^{abcd}R_{ab}^{ef}R_{cdef}$, and with
$\epsilon^{abcd}$ being the totally antisymmetric Levi-Civita
tensor. We shall assume that $f(R)$ will have the form,
\begin{equation}\label{starobinskyreview}
f(R)=R+\frac{1}{M^2}R^2\, .
\end{equation}
The equations of motion for a FRW metric are,
\begin{align}\label{eqnsofmkotion}
& 3 H^2F=\kappa^2\frac{1}{2}\dot{\phi}^2+\frac{RF-f+2
V\kappa^2}{2}-3H\dot{F}\, ,\\ \notag &
-3FH^2+2\dot{H}F=\kappa^2\frac{1}{2}\dot{\phi}^2-\frac{RF-f+2
V}{2}+\ddot{F}+2H\dot{F}\, .
\end{align}
The Chern-Simons term does not affect the background field
equations, and issue stressed also in the literature
\cite{Hwang:2005hb,Choi:1999zy}. As it was shown in
\cite{Hwang:2005hb,Choi:1999zy} only the tensor perturbations are
affected by the Chern-Simons coupling, while the scalar
perturbations remain unaffected. It can be shown that the
tensor-to-scalar ratio and the spectral index are,
\cite{Odintsov:2019mlf},
\begin{equation}\label{spectralstarobinsky}
n_s=1-\frac{2}{N},\,\,\,r\simeq
\frac{r_s^v}{2}\left(\frac{1}{|1-\frac{\kappa^2x}{F}|}+\frac{1}{|1+\frac{\kappa^2x}{F}|}\right)\,
.
\end{equation}
with $r_s^v= 48\epsilon_1^2$ being the tensor-to-scalar ratio of
the vacuum $f(R)$ gravity. The effect of the Chern-Simons term is
quantified by the presence of the term $\sim \frac{\kappa^2x}{F}$
in the tensor-to-scalar ratio. The value of the tensor-to-scalar
ratio for the vacuum $f(R)$ gravity case is $r_s^v=0.0033$,
however the Chern-Simons induced term can further reduce the value
of the tensor-to-scalar ratio, for example if
$\frac{\kappa^2x}{F}=\mathcal{O}(3\times 10^{8})$, the
tensor-to-scalar ratio becomes $r=\mathcal{O}(10^{-11})$. Thus we
demonstrated that the tensor-to-scalar ratio of the $R^2$ model
can be further reduced by the Chern-Simons term.

\section{Brief Discussion on Future Perspectives: Neutron Stars and Gravitational Waves}

In this compact review we studied the phenomenology of axion
$F(R)$ gravity models. What remains to be done in this type of
theories is to appropriately study the reheating era. During that
era we expect the $F(R)$ gravity to dominate this era, since the
axion scalar scales as a dark matter particle. In addition,
another non-trivial study is related to the matter curvature
perturbations. Both the axion field and the $F(R)$ gravity have a
non-trivial effect on the matter curvature perturbations, and this
study is very relevant and necessary since the growth index is
strongly related to matter curvature perturbations. A vital study
and very timely is related to calculating the modified gravity
effects on neutron stars \cite{Astashenok:2020cfv}. As was shown
in \cite{Astashenok:2020cfv}, the axion field can affect the
maximum allowed mass for the neutron star. In addition, in
\cite{Nojiri:2019nar}, it was shown that the axion field causes
inequivalent propagation in the circularly polarized gravitational
wave modes. Both the neutron star and gravity waves issues are
timely, and perhaps should be further be studied in the axion
perspective.

\section*{Acknowledgments}

This work is supported by MINECO (Spain), FIS2016-76363-P, and by
project 2017 SGR247 (AGAUR, Catalonia) (S.D.O), and by Russian
Ministry of Science and High Education, project No. 3.1386.2017.

\end{document}